\documentclass[sigconf]{acmart}

\copyrightyear{2026}
\acmYear{2026}
\setcopyright{cc}
\setcctype{by}
\acmConference[MSR '26]{23rd International Conference on Mining Software Repositories}{April 13--14, 2026}{Rio de Janeiro, Brazil}
\acmBooktitle{23rd International Conference on Mining Software Repositories (MSR '26), April 13--14, 2026, Rio de Janeiro, Brazil}
\acmPrice{}
\acmDOI{10.1145/3793302.3793365}
\acmISBN{979-8-4007-2474-9/2026/04}

\usepackage{listings}
\usepackage{algorithmic}
\usepackage{textcomp}
\usepackage{multirow}

\usepackage{graphicx}
\usepackage{xcolor}
\usepackage{framed}

\definecolor{shadecolor}{gray}{0.93}
\definecolor{RQbar}{RGB}{0,70,140}        %
\definecolor{Findingbar}{RGB}{85, 95, 0}    %

\begin{document}

\title{Analyzing Dependency Distribution Changes Arising from Code Smell Interactions}

\author{Zushuai Zhang}
\affiliation{%
  \institution{University of Auckland}
  \city{Auckland}
  \country{New Zealand}
}
\email{derek.zhang@auckland.ac.nz}

\author{Elliott Wen}
\affiliation{%
  \institution{University of Auckland}
  \city{Auckland}
  \country{New Zealand}
}
\email{elliott.wen@auckland.ac.nz}

\author{Ewan Tempero}
\affiliation{%
  \institution{University of Auckland}
  \city{Auckland}
  \country{New Zealand}
}
\email{e.tempero@auckland.ac.nz}

\renewcommand{\shortauthors}{Zhang, Wen, and Tempero}

\begin{abstract}
Dependencies between modules can trigger ripple effects when changes are made, making maintenance complex and costly, so minimizing these dependencies is crucial. Consequently, understanding what drives dependencies is important. One potential factor is code smells, which are symptoms in code that indicate design issues and reduce code quality. When multiple code smells interact through static dependencies, their combined impact on quality can be even more severe. While individual code smells have been widely studied, the influence of their interactions remains underexplored. In this study, we aim to investigate whether and how the distribution of static dependencies changes in the presence of code smell interactions. We conducted a dependency analysis on 116 open-source Java systems to quantify these interactions by comparing cases where code smell interactions exist and where they do not. Our results suggest that overall, code smell interactions are linked to a significant increase in total dependencies in 28 out of 36 cases, and that all code smells are associated with a consistent change direction (increase or decrease) in certain dependency types when interacting with other code smells. Consequently, this information can be used to support more accurate code smell detection and prioritization, as well as to develop more effective refactoring strategies.

\end{abstract}



\maketitle

\section{Introduction}

Modularity is a critical aspect of software maintainability. When excessive dependencies accumulate between modules, even small changes can trigger ripple effects across the codebase, making maintenance activities complex and costly \cite{cerny2024maintainability}. Practitioners therefore stress reducing inter-module dependencies to maintain high modularity. However, our understanding of which characteristics of code drive dependencies remains limited. In this paper, we investigate a prominent suspect: code smells, examining whether the number and nature of dependencies change in the presence of multiple code smells.

Code smells are characteristics of source code that indicate potential design flaws \cite{fowler1999refactoring} and are believed to degrade maintainability \cite{fowler1999refactoring, munro2005product, marinescu2012assessing, yamashita2013code, soh2016code, singh2019reducing}. Interestingly, studies of isolated code smells often report non-significant impacts \cite{sjoberg2012quantifying, yamashita2014assessing}, while those on multiple code smells frequently find significant effects \cite{singh2019reducing, palomba2018diffuseness, santana2024unraveling}. This pattern, along with other studies \cite{pietrzak2006leveraging, lanza2007object, moha2009decor, yamashita2013exploring, yamashita2015inter}, suggests that when multiple code smells are present, their interactions through static dependencies may play a more critical role in degrading maintainability, further complicating maintenance and refactoring. However, to the best of our knowledge, few studies empirically investigate the impact of such interactions on software quality \cite{yamashita2013exploring, yamashita2015inter}, and none analyze their influence on dependency growth.

The primary objective of this study is to investigate whether code smell interactions are associated with growth in the number of overall and specific types of static dependencies. To achieve this, we analyze 116 top-starred open-source Java systems from GitHub, focusing on five commonly studied code smells: Feature Envy, Brain Method, God Class, Brain Class, and Data Class. We compare the number of static dependencies in situations where code smell interactions exist and where they do not. The analysis covers both the total number of dependencies and the counts for each specific dependency type. Non-parametric statistical tests, including the Mann-Whitney U test and Cliff’s Delta, are applied to evaluate statistical significance and effect size.

In summary, our contributions are as follows:

\begin{itemize}

    \item We designed and conducted a large-scale empirical analysis involving 116 Java systems and five types of code smells. All source code, data (e.g., code smell detection output, metric measurements for code smell detection, dependency detection output, and statistical analysis results), and replication instructions are available in our replication package \cite{blind_2025_15624918}.

    \item We provide the first large-scale empirical evidence that code smell interactions are associated with an increasing number of dependencies, along with consistent change directions (increase or decrease) for certain dependency types. Our results offer actionable guidance for code smell prioritization and detection, as well as for developing more effective refactoring strategies, benefiting tool developers and practitioners aiming to improve software maintainability by removing code smells and decoupling modules. Researchers can also build on our study design and findings on changes in dependency distribution to further investigate code smell interactions.

\end{itemize}

The remainder of this paper is organized as follows. Section~\ref{sec:background} presents the background and related work. Section~\ref{sec:study_design} describes the study design, including our research questions and implementation details. Section~\ref{sec:results} presents the results and answers the research questions, while Section~\ref{sec:discussion} discusses the practical implications of the results and potential threats to validity. Finally, Section~\ref{sec:conclusions} provides the conclusions and outlines directions for future work.

\section{Background and Related Work} \label{sec:background}

\subsection{Code Smells}
\label{sec:background_detection}

The term \textit{code smell} was introduced by Kent Beck in the late 1990s \cite{bansal2022categorical} and gained popularity after Martin Fowler described 22 of them \cite{fowler1999refactoring}. Code smells can be classified into method-level and class-level categories, as some affect the maintainability of a single method, while others influence the maintainability of an entire class \cite{lanza2007object}.

In this study, we focus on the following five \cite{lanza2007object}:

\begin{itemize}
  \item \textbf{Feature Envy (FE):} A method that relies more on another class’s data than on its own.
  \item \textbf{Brain Method (BM):} An overly large, complex, and deeply nested method that centralizes too much functionality within a single class.
  \item \textbf{God Class (GC):} A class that centralizes the system’s functionality, relies heavily on data from other classes, and exhibits low cohesion, indicating multiple responsibilities.
  \item \textbf{Brain Class (BC):} An excessively large class that concentrates too much of the system’s functionality, typically containing at least one Brain Method. It has low cohesion but can be more cohesive than a God Class.
  \item \textbf{Data Class (DC):} A low-complexity class that primarily holds data and exposes it through public fields or accessor methods instead of encapsulating behavior.
\end{itemize}

We selected these smells because (i) they represent both class-level and method-level design flaws, (ii) they are among the most frequently considered and empirically investigated code smells in the software maintenance literature \cite{kaur2020systematic}, (iii) they have often been empirically identified as among the most prevalent in codebases \cite{reis2020code, zakeri2023systematic} and harmful to maintainability \cite{li2007empirical, cairo2018impact, tufano2017and, palomba2018diffuseness}, and (iv) Lanza and Marinescu \cite{marinescu2005measurement, lanza2007object} proposed a well-justified and explainable detection approach for them, called the \textit{metric-based approach}, which uses a set of metrics with predefined thresholds to identify code smell symptoms. This approach has been adopted by many popular automatic detection tools, such as PMD\footnote{\url{https://pmd.github.io/}}.

\subsection{Influence of Code Smells on Maintainability}
\label{subsec:interelations_interactions}

Code smells are widely believed and have been empirically shown to compromise software maintainability \cite{fowler1999refactoring, munro2005product, lozano2008assessing, olbrich2010all, d2010impact, sjoberg2012quantifying, marinescu2012assessing, yamashita2013code, sabane2013study, chatterji2013effects, ban2014recognizing, yamashita2014assessing, soh2016code, rahman2017relationships, mondal2017bug, zhang2017empirical, islam2017comparative, ban2017connection, jaafar2017analyzing, elish2017association, mondal2018cloned, barbour2018investigation, palomba2018diffuseness, singh2019reducing, aversano2020empirical, imran2022qualitative, masudur2025software}. A key observation is that code smells rarely exist in isolation; instead, multiple code smells often \emph{co-locate} within the same class and collectively exacerbate maintainability issues \cite{pietrzak2006leveraging, yamashita2015inter, abbes2011empirical, yamashita2013exploring, martins2020code, martins2021code, hamdi2021empirical, santana2024exploratory, santana2024unraveling}.

Abbes et al. \cite{abbes2011empirical} investigated the effects of the Blob and Spaghetti code smells on code comprehensibility. They reported that classes exhibiting neither smell, or only one of the two, showed no measurable impact; however, when the two smells co-located in the same class, comprehension tasks required more time and effort and resulted in lower accuracy. Hamdi et al. \cite{hamdi2021empirical}, focusing on Android applications, found that 51\% of classes in 1,923 apps contained at least two smells, with 14 specific smell pairs frequently co-locating. Martins et al. \cite{martins2021code} and Santana et al. \cite{santana2024exploratory} both found that classes containing multiple types of code smells tend to exhibit higher complexity and lower cohesion than those with only one or none, with certain combinations being particularly detrimental, such as a God Class co-located with a Feature Envy. Furthermore, the developer survey by Martins et al. \cite{martins2021code} reported that code containing co-located smells was considerably harder to understand and modify. Santana et al. \cite{santana2024unraveling} additionally showed that classes with co-located code smells undergo significantly more frequent and extensive changes, indicating reduced code stability. These findings align with earlier studies \cite{sjoberg2012quantifying, yamashita2014assessing}, which reported that individual smells alone have limited effects on maintenance effort, implying that smell co-location is a more important factor in real maintenance tasks.

Interestingly, it has been suggested that maintenance problems in large classes with co-located code smells are sometimes caused not by complexity itself, but by interactions between the smells through static dependencies \cite{yamashita2012assessing, yamashita2015inter}. Building on this idea, Yamashita and Moonen \cite{yamashita2013exploring} conducted a study involving six professional developers who performed real change tasks in four Java systems. They recorded daily interviews and think-aloud sessions. The authors observed that code smells can interact with each other through static dependencies not only when co-located, but also across different classes. Such cases were referred to as \textit{coupling}, where code smells in different classes are connected by static dependencies. Based on a qualitative analysis of developer feedback, they found that these interactions often hinder code comprehension and change propagation due to the ripple effects of static dependencies, thereby increasing maintenance effort and defect risk. To the best of our knowledge, this remains the only study that has examined the influence of such interactions on maintainability. Subsequent work has typically focused on identifying common patterns, such as which code smells tend to be collocated or coupled \cite{yamashita2015inter, fontana2015towards, walter2018code, sobrinho2021interplay, martins2020code}, but has rarely linked these findings to maintainability.

In summary, previous research has extensively examined the influence of code smell co-location on maintainability, but has rarely considered the factor of interaction. The only study addressing this issue was qualitative and based on a small dataset \cite{yamashita2013exploring}. Our work differs fundamentally by linking interaction to a maintainability metric, namely the number of static dependencies as an indicator of modularity. Using a significantly larger dataset, we apply statistical tests to enable quantitative assessment. Through a detailed analysis of each type of static dependency and its flow direction, we provide practical and actionable insights for improving modularity, guiding code smell refactoring, prioritization, and detection.

\section{Study Design} \label{sec:study_design}

\subsection{Research Questions}

This study addresses two research questions (RQs):

\begin{description}
  \item[\textbf{RQ1}] How frequently do code smells interact?
  \item[\textbf{RQ2}] How do dependency distributions differ due to code smell interactions?
\end{description}

Our main focus is RQ2, where we statistically compare the number of dependencies between situations in which code smell interactions exist and those in which they do not. However, before addressing this question, we first examine RQ1 to determine how frequently code smells interact. If such interactions are rare, investigating their consequences would be less meaningful.

\subsection{Dependency}
\label{subsec:study_design_dependency}

To answer our RQs, we need to define what we mean by ``interaction,'' and to do that, we need to define what we mean by ``artifact'' and ``dependency.'' We consider the following \textbf{\emph{artifact types}}, organized in a hierarchical structure where types at the same level are mutually exclusive: Type (including both Class and Interface), Callable (including Method and Constructor), and Field. We further classify Method into Accessor (i.e., getters and setters) and Functional Method (FM; i.e., methods that are not accessors), because calling an accessor provides access to fields, whereas calling a functional method typically executes business logic.

We define a \textbf{\emph{dependency type}} as the triple
\[
(\text{relation},\ \text{source artifact type},\ \text{target artifact type}),
\]

\noindent
where relation is the abstract kind of dependency (e.g., call), and source artifact type and target artifact type are the leaf artifact types in the hierarchy, providing more detail and avoiding double-counting (i.e., Class, Interface, Functional Method, Accessor, Constructor, and Field). Each dependency type is directional, flowing from the source artifact to the target artifact. We treat two dependency types as distinct whenever any of the three fields in the triple differ (for example, (call, FM, FM) $\neq$ (call, Constructor, FM)).

Table~\ref{tab:dependency_type} lists all dependency types considered in this study, along with their descriptions. Some relations can occur between multiple source and target artifact types; therefore, we present the possible source and target types as sets. Every ordered pair formed by taking one element from the source set and one element from the target set is valid.

\begin{table*}[tb]
    \centering
    \caption{Dependency Types}
    \Description{Dependency Types}
    \begin{tabular}{@{}p{1.1cm} p{3.7cm} p{2.9cm} p{9.0cm}@{}}
        \hline
        \textbf{Relation} & \textbf{Source Artifact Type} & \textbf{Target Artifact Type} & \textbf{Description} \\ 
        \hline
        call & \{FM, Constructor, Field\} & \{FM, Accessor, Constructor\} & An FM, constructor, or field initializer invokes an FM, accessor, or constructor. \\

        \hline
        create & \{FM, Constructor, Field\} & \{Class\} & An FM, constructor, or field initializer instantiates (i.e., uses \texttt{new}) a class. \\

        \hline
        contain & \{FM, Constructor, Field\} & \{Class\} & An FM or constructor declares a local variable of a certain class type, or a field is declared with a class type.\\ 

        \hline
        cast & \{FM, Constructor, Field\} & \{Class\} & An FM, constructor, or field initializer explicitly casts an object to a class type. \\

        \hline
        use & \{FM, Accessor, Constructor, Field\} & \{Field\} & An FM, accessor, constructor, or field initializer references a field. \\ 

        \hline
        throws & \{FM, Constructor\} & \{Class\} & An FM or constructor declares that it may throw an exception (a class) or explicitly throws one within its body. \\

        \hline
        return & \{FM, Accessor\} & \{Class\} & An FM or accessor returns an object of a class type. \\

        \hline
        parameter & \{FM, Accessor, Constructor\} & \{Class\} & An FM, accessor, or constructor has a parameter of a certain class type. \\ 

        \hline
        extend & \{Class\} & \{Class\} & A class extends another class. \\

        \hline
        implement & \{Class\} & \{Interface\} & A class implements an interface. \\
        \hline
    \end{tabular}
    \label{tab:dependency_type}
\end{table*}

\subsection{Interaction}
\label{subsec:interaction}

\begin{figure}[tb]
  \centering
\begin{lstlisting}
  class C1 {
    public void m1(C2 c2) { c2.m2(); }
  }
  class C2 {
    public void m2() { System.out.println("m2"); }
    public void m3() { new C1().m1(this); }
  }
\end{lstlisting}
  \caption{Interaction Example}~\label{fig:interaction_example}
  \Description{Interaction Example}
\end{figure}

Two artifacts, a1 and a2, interact if there exists a static dependency connecting them. The set of such dependencies constitutes the \textbf{\emph{interaction set}}; if it is empty, no interaction occurs. Any interaction has the property of \textbf{\emph{relative location}}, with a value of Same or Different, indicating whether the artifacts reside in the same or in different classes. This distinction is motivated by the observation that dependency distributions may vary with relative location (e.g., when accessing data in different classes, FM typically \textit{call} accessors, but when in the same class, FM typically \textit{use} fields).

For example, consider the interaction between the method \texttt{m1} and the class \texttt{C2}, which are from different classes, as shown in Figure~\ref{fig:interaction_example}. The method \texttt{m1} takes a \texttt{C2} object as a parameter and calls \texttt{C2\#m2} through that object. Additionally, \texttt{C2\#m3} creates a \texttt{C1} object and invokes \texttt{m1}. The interaction set consists of three dependencies: (i) a (parameter, FM, Class) dependency from \texttt{m1} to \texttt{C2}, (ii) a (call, FM, FM) dependency from \texttt{m1} to \texttt{C2\#m2}, and (iii) a (call, FM, FM) dependency from \texttt{C2\#m3} to \texttt{m1}. The (create, FM, Class) dependency from \texttt{C2\#m3} to \texttt{C1} is not in the set because it does not link \texttt{m1} and \texttt{C2}. It would be included if we considered the interaction between \texttt{C1} and \texttt{C2} or between \texttt{C1} and \texttt{C2\#m3}.

\subsection{Code Smell Interaction}

Let CS1 and CS2 be two code smell types, and let cs1 and cs2 be artifacts that exhibit CS1 and CS2, respectively. For any smell type CS, we denote by nonCS its non-code smell instance: an artifact of the same type that does not exhibit CS (though it may exhibit other smells). For example, a non-God Class (nonGC) is a class that is not a GC, although it may exhibit other smells (e.g., DC). If cs1 interacts with cs2, we say there exists a code smell interaction for the pair (CS1, CS2). We use the shorthand CS1-CS2 to denote all such cases.

Conversely, if cs1 interacts with a nonCS2 artifact, this represents a case where a code smell interaction for the pair (CS1, CS2) does \emph{not} exist, because the second artifact lacks CS2. We denote such cases by CS1-nonCS2. Symmetrically, nonCS1-CS2 denotes cases where a code smell interaction for the pair (CS1, CS2) does not exist because the first artifact lacks CS1. We refer to these three situations collectively as \textbf{\emph{interaction types}}: one with a code smell interaction (CS1-CS2) and two without (CS1-nonCS2, nonCS1-CS2).

Because dependencies are directional, each dependency in the interaction set is assigned a \textbf{\emph{flow direction}} to relate direction and code smell context. The order of the two code smells is not inherently significant; however, once fixed (e.g., in our example, the order is set to (CS1, CS2)), it determines how the flow direction is defined:

\begin{itemize}
    \item \textbf{Forward}: The CS1 or nonCS1 artifact is the source, and the CS2 or nonCS2 artifact is the target.
    
    \item \textbf{Backward}: The CS2 or nonCS2 artifact is the source, and the CS1 or nonCS1 artifact is the target.
\end{itemize}

We use ``$\rightarrow$'' to indicate that only dependencies in a specific direction are considered. For example, \textbf{CS2$\rightarrow$CS1} means that we consider only those dependencies where the CS2 artifact is the source and the CS1 artifact is the target (i.e., \emph{backward}).

\subsection{Data Collection}

\begin{figure}[htb]
\centering
  \includegraphics[width=0.8\columnwidth]{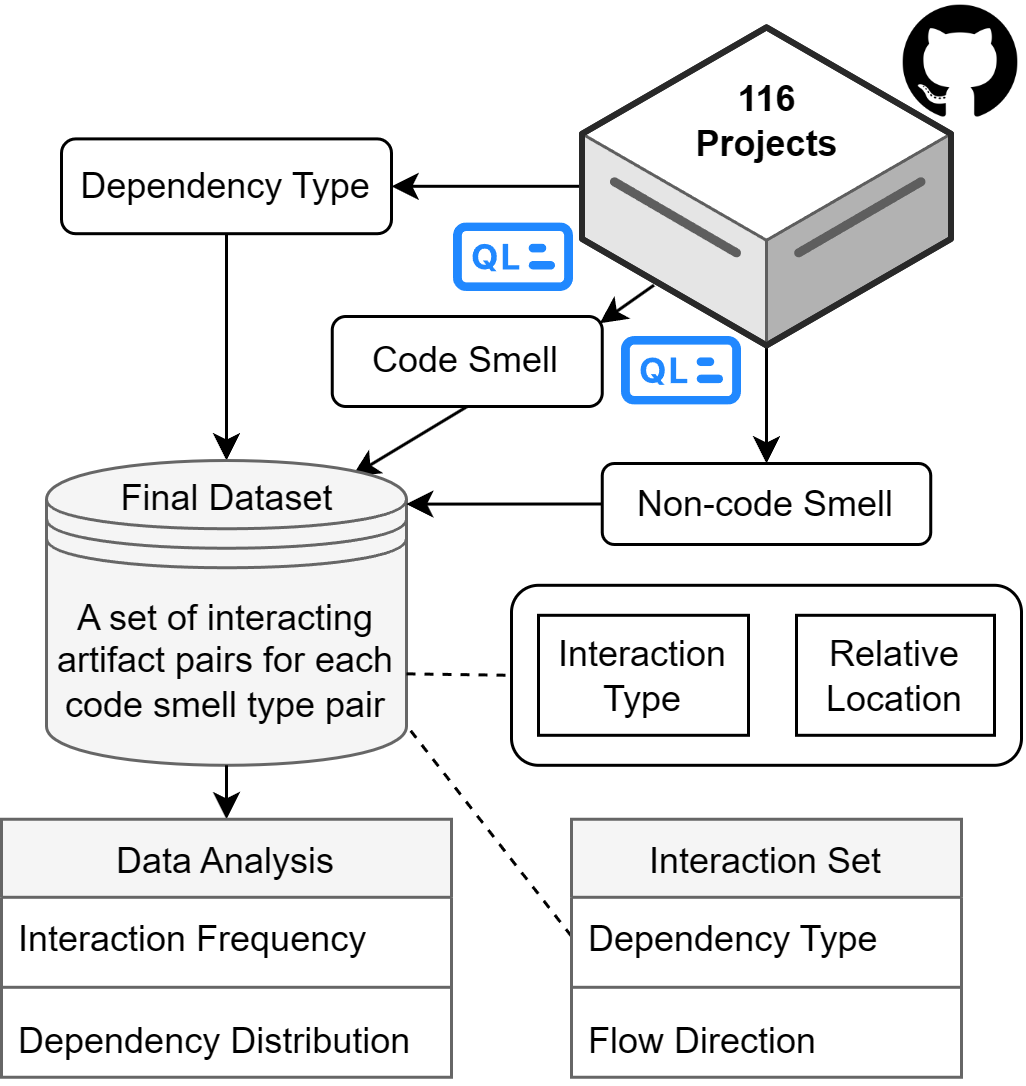}
  \caption{Data Collection Overview}~\label{fig:data_collection}
  \Description{Data Collection Overview}
\end{figure}

Figure~\ref{fig:data_collection} provides an overview of data collection. Arrows indicate the direction of data used as input for the next step. Dashed lines indicate inspecting the inside of the data component. A total of 116 Java projects were selected from GitHub (see Section~\ref{sec:system_selection}). A tool was implemented using CodeQL\footnote{https://codeql.github.com/} (see Section~\ref{sec:implementation}) to extract all dependency types, artifacts of five code smells, and their corresponding non-code smells. After combining these three pieces of information, for each pair of code smell types (CS1, CS2), we obtain a set of interacting artifact pairs (e.g., \{(a1, a2), (a3, a4), ...\}), each corresponding to one of the three interaction types for (CS1, CS2). The collection of all such sets across every code smell type pair constitutes the final dataset, which is then used for data analysis (see Section~\ref{sec:data_analysis}). For each interacting artifact pair in the final dataset, we record its interaction type and relative location. The interaction set is also recorded with dependency type and flow direction information for each of the dependency instances within the set.

\subsection{Data Analysis} \label{sec:data_analysis}

To answer RQ1, data from all systems were combined (hereafter referred to as the \textbf{\emph{Union dataset}}). For each code smell pair, CS1 and CS2, \textbf{\emph{interaction frequency}} is calculated, defined as the percentage of distinct CS1 artifacts that interact with at least one CS2 artifact (i.e., with a non-empty interaction set) relative to the total number of CS1 occurrences. The same procedure was then applied from the perspective of CS2.

For RQ2, to determine whether dependency distributions vary due to code smell interactions, for a code smell pair (CS1, CS2), we statistically test whether the number of dependencies differs between interaction types where code smell interactions exist and interaction types where they do not, with relative location as a moderator (i.e., independent variable: interaction type; moderator variable: relative location). Therefore, we test four \textbf{\emph{contrasts}} (the notations on the right indicate the shorthand labels used for each contrast):

\begin{itemize}
    \item (CS1-CS2) vs. (CS1-nonCS2), Different \hfill (\textit{nonCS2, Diff.})
    \item (CS1-CS2) vs. (nonCS1-CS2), Different \hfill (\textit{nonCS1, Diff.})
    \item (CS1-CS2) vs. (CS1-nonCS2), Same \hfill (\textit{nonCS2, Same})
    \item (CS1-CS2) vs. (nonCS1-CS2), Same \hfill (\textit{nonCS1, Same})
\end{itemize}

These four contrasts are tested on two kinds of dependent variables: (i) General analysis: the total number of dependencies between the two artifacts (i.e., the interaction set size); and (ii) Specific analysis: the number of dependencies of a specific dependency type and flow direction within the interaction set.

The testing was conducted from two analytical \textbf{\emph{perspectives}}: (i) Union: using the Union dataset to provide an overall overview and identify general trends; and (ii) Per-system: testing each system's data individually to determine whether code smell interactions vary across systems, i.e., whether the general trend holds within each system.

\noindent\textbf{Hypotheses.}
For a code smell pair $(\mathrm{CS1},\mathrm{CS2})$, let $t_1$ and $t_2$ be interaction types, where
$t_1=\mathrm{CS1\text{-}CS2}$ and
$t_2\in$ \{CS1-nonCS2,nonCS1-CS2\}.
Let $l\in\{\mathrm{Different},\mathrm{Same}\}$ denote the relative location,
$d\in\{\mathrm{General\ analysis},\mathrm{Specific\ analysis}\}$ denote the dependent variable of interest,
and $p\in\{\mathrm{Union},\text{Per-system}\}$ denote the analysis perspective. To answer RQ2, we test the following hypotheses:

\vspace{0.5\baselineskip}

{%
\small\noindent
{\renewcommand{\arraystretch}{1.2}%
\begin{tabular}{@{}p{1.4cm}p{\dimexpr\linewidth-1.4cm-2\tabcolsep\relax}@{}}
$\mathbf{H_{0\text{-}pldt_1t_2}:}$ &
Under perspective $p$ and location $l$, the distribution of $d$ for inter\-action types $t_1$ and $t_2$ are equal. \\[.25ex]
$\mathbf{H_{1\text{-}pldt_1t_2}:}$ &
Under perspective $p$ and location $l$, the distribution of $d$ for inter\-action types $t_1$ and $t_2$ are not equal.
\end{tabular}}
}

The Mann-Whitney U test \cite{mann1947test} was selected as the primary statistical test. This test was chosen because our data does not follow a normal distribution, and the two interaction types being compared are independent with unequal sample sizes (e.g., there are more interacting artifact pairs that are CS1-nonCS2 or nonCS1-CS2 than CS1-CS2, due to the larger number of non-code smells compared to code smells).

Since this test only determines whether a difference exists but does not measure its magnitude, we computed Cliff’s Delta ($\delta$), a non-parametric effect size measure that quantifies the strength of the observed difference. Cliff’s Delta can be positive or negative. A positive value indicates that code smell interactions are associated with an increased number of dependencies, while a negative value suggests a decrease. To interpret Cliff’s Delta, we adopted widely accepted thresholds \cite{vargha2000critique}: negligible ($|\delta| < 0.147$), small ($0.147 \leq |\delta| < 0.33$), medium ($0.33 \leq |\delta| < 0.474$), and large ($|\delta| \geq 0.474$). We refer to the sign of Cliff’s Delta as the \textbf{\emph{change direction}}, indicating whether it represents an increase or a decrease, and to the absolute value as the \textbf{\emph{change magnitude}}, reflecting the strength of the difference.

Certain interaction types at specific relative locations for some code smell pairs are unlikely to occur due to the inherent nature of the code smells. For example, our dataset does not contain any class that is simultaneously a BC or GC and a DC, because BC and GC are complex classes, whereas DC are simple classes. Additionally, certain dependency types are impossible for specific flow directions within certain code smell pairs due to differences in granularity. For instance, a (use, FM, Field) dependency cannot occur from GC to BM, as BM is a method-level smell and therefore cannot serve as a field type target artifact. Furthermore, some projects may not contain certain code smells or may contain them without any interactions. In such cases, statistical tests cannot be performed. However, the conclusions of this study remain valid, as we focus only on cases where interactions occur.

\noindent
\textbf{Multiple Testing Correction}. To reduce the risk of false discoveries, we applied the Benjamini-Hochberg procedure in our analyses. For the Union dataset, we adjusted across 10 (pairs) $\times$ 4 (contrasts) $ = 40$ p-values for the general analysis, and 10 $\times$ 4 $\times$ 2 (flow directions) $\times$ 31 (dependency types; considering all possible triples in Table~\ref{tab:dependency_type}) $ = 2{,}480$ p-values for the specific analysis. For the per-system dataset, we adjusted across $40\times 116$ (number of systems)$ = 4{,}640$ p-values for the general analysis, and $2{,}480 \times 116=287{,}680$ p-values for the specific analysis. Note that these numbers represent upper bounds; the actual number of p-values is lower because some cases are unlikely to occur, as explained in the previous paragraph.

\subsection{Systems Selection}
\label{sec:system_selection}

We focus on open-source Java projects on GitHub. The rationale for assessing only Java projects is to ensure consistency with previous research for comparable results. To prioritize the most popular and widely accepted projects, we sorted all available projects in descending order based on the number of stars. Then, we applied the following inclusion and exclusion criteria (IC and EC):
\textbf{IC1}. More than 100 forks (indicating active use and acceptance by the community).
\textbf{IC2}. Java Percentage: Over 80\% of the source code must be written in Java (due to tool limitations).
\textbf{EC1}. Educational projects that do not represent real-world software development practices.
\textbf{EC2}. Projects that are no longer maintained.

We selected the first 150 projects meeting the ICs and manually analyzed them to apply the ECs, resulting in 116 retained projects. Table \ref{tab:proj_summary_stat} provides a statistical summary of the 116 Java projects analyzed, covering key software metrics such as the number of classes (NOC), number of methods (NOM), lines of code (LOC, excluding blank lines and comments), and the occurrence of five code smells. The dataset varies greatly in size: the standard deviation (SD) of LOC is almost twice the mean. The largest project (Hadoop) contains over 2.7 million lines of code, whereas the smallest (RxAndroid) has just over 1,300. Similarly, the number of code smells varies from zero to high occurrences. Given the diversity in project scale and code smells, we believe this sample is representative of real-world Java software and provides sufficient data for analyzing code smell interactions.

\begin{table}[tb]
    \centering
    \caption{Projects Summary Statistics}
    \Description{Projects Summary Statistics}
    \begin{tabular}{lllllll}
        \hline
        & Mean & Median & Max & Min & Sum & SD \\
        \hline
        NOC  & 2.8K  & 1.1K    & 22.1K   & 15  & 321.8K & 4.2K \\
        NOM  & 24.5K & 8.5K  & 338.7K  & 124 & 2.8M & 45.3K \\
        LOC  & 234.2K & 78.9K & 2.7M & 1.3K & 27.2M & 421.9K \\
        \#BC & 6.4     & 1       & 101     & 0   & 737 & 14.7   \\
        \#GC & 20.2    & 6       & 209     & 0   & 2.4K & 34.2    \\
        \#BM & 26.7    & 5       & 294     & 0   & 3.1K & 54.7    \\
        \#FE & 30.5    & 9       & 275     & 0   & 3.5K & 51.6    \\
        \#DC & 45.3    & 15      & 384     & 0   & 5.3K & 66.2    \\
        \hline
        \multicolumn{7}{l}{K = thousand; M = million; SD = standard deviation} \\
    \end{tabular}
    \label{tab:proj_summary_stat}
\end{table}

\subsection{Implementation} \label{sec:implementation}

CodeQL\footnote{https://codeql.github.com/} is a static analysis tool originally developed to identify vulnerabilities in codebases. Beyond security analysis, CodeQL allows users to detect custom code patterns using a declarative query language~\cite{youn2023declarative}. In this study, it was employed to detect code smells, analyze dependencies, and examine interactions. We employed the popular and widely used metric-based approach proposed by Lanza and Marinescu~\cite{lanza2007object} to detect code smells, as described in Section~\ref{sec:background_detection}. Our detection queries were implemented using the metrics and thresholds they provided (see Section \ref{subsec:threat} for why we built our own smell detector). When a threshold was specified as a range, we selected the stricter bound to minimize the false positive rate. That is, for metrics required to be smaller than the threshold, we selected the lower bound, and for metrics required to be larger than the threshold, we selected the upper bound. For each project, we excluded third-party libraries from analysis, as developers typically do not have control over their source code. Our ultimate goal is to ensure that our findings are actionable for developers. Additionally, we excluded test and example code and focused exclusively on production code. See our replication package for details on code smell and metric definitions, as well as the chosen thresholds and detection rules~\cite{blind_2025_15624918}.

\noindent
\textbf{Evaluation}. To validate detection correctness, we computed the per-smell sample size using the standard confidence interval for a proportion (95\% confidence, $\pm 10\%$ margin) and then performed \emph{stratified proportional sampling by project}: for each smell, we allocated $n_i \propto N_i$ (where $N_i$ is the number of detections in project $i$) and drew simple random samples within each project \cite{cochran1977sampling}. This yields an unbiased, lower-variance estimate of the overall false positive rate while ensuring that representation is aligned with each project's contribution \cite{singh1996stratified}. The first author then performed a manual evaluation of the sampled smell instances, first checking whether the metric measurements were correct and then checking whether each detected instance was strictly aligned with the definition of that smell provided by Lanza and Marinescu~\cite{lanza2007object}. Our tool also outputs information related to each metric (e.g., for metrics used to detect FE, it reports which foreign fields are accessed and from which classes), which allows us to quickly verify the measurements by checking whether this information is correct. In total, we manually checked 94, 94, 93, 86, and 95 samples for FE, BM, GC, BC, and DC, respectively. All sampled detected smell instances and their metric measurements were judged correct (100\%), with no false positives observed in the sample, indicating that these instances are consistent with the definitions we used. We also applied the same procedure to non-code smells by manually checking 97, 97, 96, 96, and 96 samples for nonFE, nonBM, nonGC, nonBC, and nonDC, respectively. All metric measurements were judged correct, and the false negative rates were low: 2.1\%, 6.2\%, 9.3\%, 8.3\%, and 11.5\%, respectively.

\noindent
\textbf{Replication.} All analyses were executed on a desktop machine (Windows 10, AMD Ryzen 9 5950X, 32 GB RAM). To replicate the results, run the provided CodeQL queries on the pre-built CodeQL databases and execute the accompanying Python scripts. A Java runtime is required only to build CodeQL databases; because we provide pre-built databases, Java is not required to replicate our results. On our machine, the full pipeline completed in approximately 5 hours. Runtime may vary with CPU performance (e.g., core count and clock speed) and storage performance. Detailed replication instructions are included in the replication package~\cite{blind_2025_15624918}.

\section{Results} \label{sec:results}

This section presents our results. Due to space constraints, we include only the most interesting results. Our replication package provides the complete dataset \cite{blind_2025_15624918}.

\subsection{Interaction Frequency}

Table \ref{tab:interaction_frequency} presents the interaction frequency in the Union dataset. The row and column headers include the code smell names and their total occurrences. Each cell displays the percentage and absolute number of distinct code smell artifacts from the row header that interact with at least one artifact from the corresponding column header. For example, the cell at row BC and column GC indicates that 79.8\% of BC artifacts interact with at least one GC artifact, with an absolute count of 588 out of a total of 737 BC artifacts.

\begin{table}[tb]
    \centering
    \caption{Interaction frequency with absolute numbers}
    \Description{Interaction frequency with absolute numbers}
    \renewcommand{\arraystretch}{1.5}
    \begin{tabular}{|p{0.7cm}|p{0.7cm}|p{0.7cm}|p{0.7cm}|p{0.7cm}|p{0.7cm}|}
        \hline
        & \textbf{BC (737)} & \textbf{GC (2348)} & \textbf{BM (3093)} & \textbf{FE (3540)} & \textbf{DC (5260)} \\
        \hline
        \textbf{BC (737)} &  & 79.8\% (588) & 99.7\% (735) & 33.1\% (244) & 44.8\% (330) \\
        \hline
        \textbf{GC (2348)} & 58.7\% (1378) &  & 46.8\% (1100) & 39.1\% (918) & 56.4\% (1324) \\
        \hline
        \textbf{BM (3093)} & 56.7\% (1753) & 49.6\% (1533) &  & 7.0\% (217) & 19.8\% (613) \\
        \hline
        \textbf{FE (3540)} & 18.0\% (637) & 46.1\% (1631) & 8.0\% (282) &  & 46.2\% (1634) \\
        \hline
        \textbf{DC (5260)} & 16.1\% (847) & 34.0\% (1791) & 11.6\% (610) & 23.1\% (1213) &  \\
        \hline
    \end{tabular}
    \label{tab:interaction_frequency}
\end{table}

Among the 20 cells in the table, around 50\% have an interaction percentage greater than 40\%, while around 80\% have an interaction percentage greater than 20\%. The highest interaction percentage occurs between BC and BM (99.7\%), but the reverse side from BM to BC is much lower (56.7\%). The lowest interactions are observed between FE and BM (8.0\%) and between BM and FE (7.0\%). GC interacts with BM (46.8\%) and FE (39.1\%) at moderate levels. A large percentage of code smells interact with GC and DC, ranging from 19.8\% to 79.8\%.

The high interaction percentage from BC to BM (99.7\%) and the much lower percentage from BM to BC (56.7\%) are because a class must contain at least one BM instance to be detected as BC. Any intra-class dependency involving the BM will be considered an interaction. However, the reverse is not necessarily true because BM can appear in a nonBC class. Separately, the low interaction percentage between BM and FE (7.0\%) occurs because both are method-level smells. The only way they can interact is through method calls. However, BM and FE both require data but do not provide it. Since calling each other does not satisfy their data requirements, they are unlikely to interact. Another observation is the high interaction percentage between BC and GC (79.8\%), possibly because they are both complex classes that depend on each other's methods.

The DC column shows a relatively high interaction frequency from all other smells to DC, with the lowest being 19.8\%. This can be explained by the fact that DC provides data, while other code smells either require data (BM, GC, and FE) or contain methods that require data (BC), making interaction likely. In contrast, when looking at the row for DC, the interaction percentages are lower than those in the column. This is because DC has the highest occurrence across 116 systems. Although many other smells obtain data from DC, the number of other smells is significantly lower than the number of DC occurrences. As a result, the interaction percentage between DC and any other smell is diluted by the large number of DC instances.

{%
\def\PadX{12pt}%
\def\PadY{4.5pt}%
\setlength{\fboxsep}{0pt}%

\def\FrameCommand#1{%
  {\color{RQbar}\vrule width 2pt}%
  \colorbox{shadecolor}{%
    \hspace*{\PadX}%
    \vbox{\kern\PadY #1 \kern\PadY}%
    \hspace*{\PadX}%
  }%
}%
\MakeFramed{\advance\hsize-\width \FrameRestore}%
\small\noindent
\textbf{RQ1 Summary.} Overall, most code smell pairs frequently interact with one another, with over 50\% of the pairs exhibiting an interaction frequency above 40\%. This highlights the importance of evaluating the consequences of code smell interactions.
\endMakeFramed
}

\subsection{General Analysis Union Dataset}

\begin{figure}[tb]
\centering
  \includegraphics[width=0.9\columnwidth]{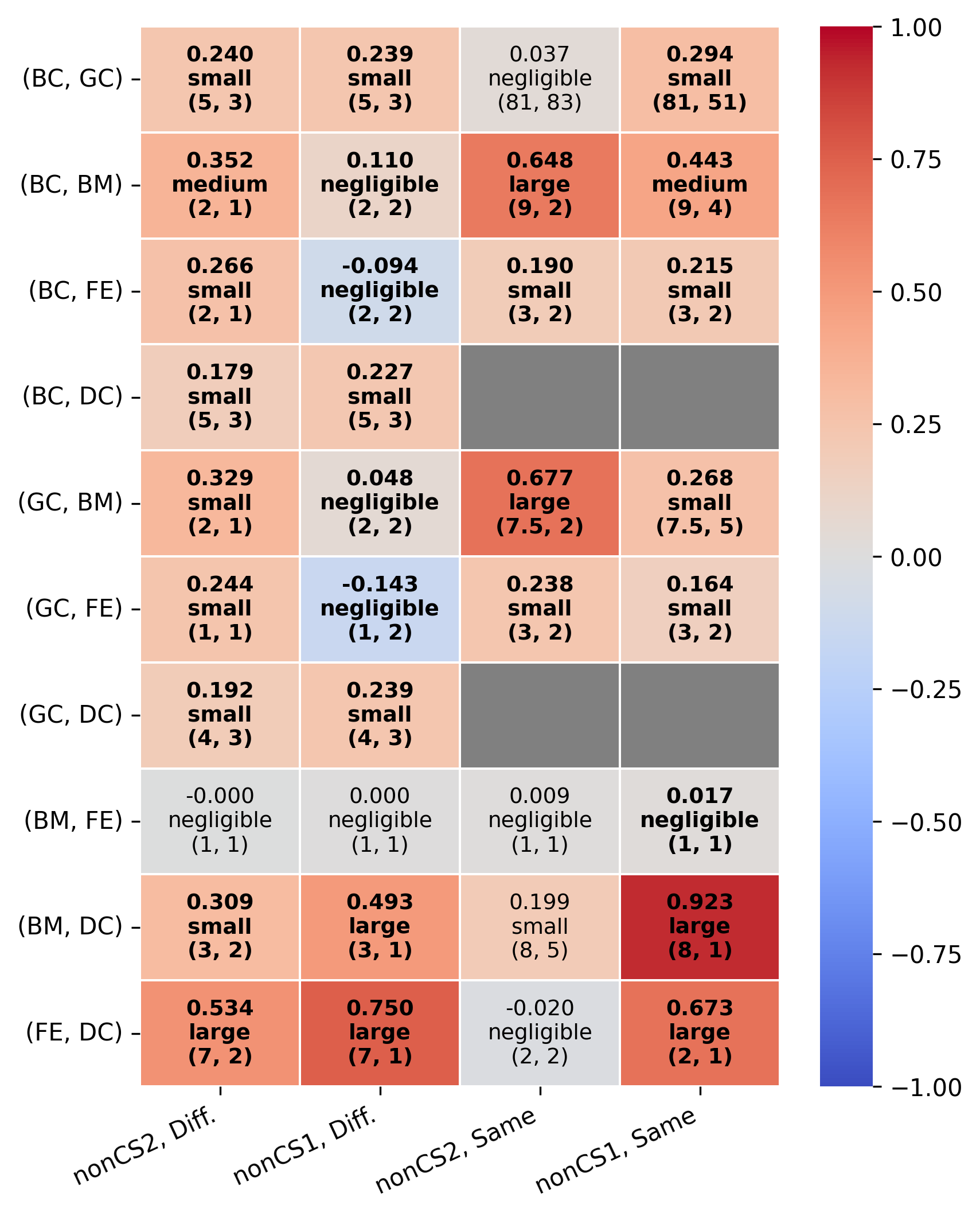}
  \caption{Cliff’s Delta Heatmap (\textbf{bold} = significant Mann-Whitney U test)}~\label{fig:heat_map}
  \Description{Cliff’s Delta Heatmap (\textbf{bold} = significant Mann-Whitney U test)}
\end{figure}

Figure~\ref{fig:heat_map} presents our general analysis results for the Union dataset (a version of this figure appears in our Doctoral Symposium paper~\cite{zhang2025theinfluence}). The y-axis shows the 10 code smell pairs, while the x-axis shows the four contrasts. Each cell contains three pieces of information: Cliff’s delta, its interpretation, and the median value for the two interaction types being compared (in parentheses). The gradient scale of the cell color is based on Cliff’s delta, ranging from red (1) to grey (0) to blue (-1). Cells in bold indicate a significant Mann-Whitney U test, where we can reject $H_0$.

Four cells are masked in gray because no class is both a BC and a DC or both a GC and a DC across the 116 projects. The entire BM-FE row exhibits a negligible effect size. Both smells occur at the method level and can interact only through method calls; therefore, a larger effect size is unlikely to be observed without examining the direction of the calls. Overall, as indicated by the heatmap's skew toward the red direction, Cliff’s delta values tend to be positive, with small to medium effect sizes. Most tests are significant, particularly in the first two columns where the two artifacts belong to different classes. Interestingly, the two cells corresponding to the (BC, FE) and (GC, FE) rows in the second column show a negligible yet statistically significant decrease.

{%
\def\PadX{12pt}%
\def\PadY{4.5pt}%
\setlength{\fboxsep}{0pt}%

\def\FrameCommand#1{%
  {\color{Findingbar}\vrule width 2pt}%
  \colorbox{shadecolor}{%
    \hspace*{\PadX}%
    \vbox{\kern\PadY #1 \kern\PadY}%
    \hspace*{\PadX}%
  }%
}%
\MakeFramed{\advance\hsize-\width \FrameRestore}%
\small\noindent
\textbf{Finding 1.} Code smell interactions are associated with a small to medium, statistically significant increase in the total number of dependencies in 28 out of 36 possible cases.
\endMakeFramed
}

\noindent
\textbf{Sensitivity Analysis.} By definition, a nonCS1 artifact may still exhibit other code smells such as CS2, which could introduce confounding effects and compromise the interpretability of the results. We conducted a sensitivity analysis by excluding nonCS artifacts that exhibit any of the other four smells (e.g., a nonCS1 artifact is already free of its own smell type, CS1, so only the other four need to be checked). The output confirms that the percentage of nonCS artifacts that are not completely clean is very low: 5.56\% for nonBC, 0.31\% for nonBM, 2.14\% for nonDC, 0.27\% for nonFE, and 4.37\% for nonGC. We reran the general analysis on the union dataset using this clean subset and regenerated Figure~\ref{fig:heat_map}. The results indicate that the effect sizes for most cells increased slightly, and a few contrasts that were previously not significant became significant. Therefore, excluding nonCS instances that are not completely clean does not invalidate our claims; rather, it further strengthens them.

\subsection{Specific Analysis Union Dataset}

\begin{table}[tb]
    \centering
    \caption{Count of tests rejecting $H_0$ across dependency types, with and without a small-effect-size threshold}
    \Description{Count of tests rejecting $H_0$ across dependency types, with and without a small-effect-size threshold}
    \begin{tabular}{|l|c|c|}
        \hline
        \textbf{Dependency Type} & \textbf{$p < .05$ and $\ge$ small effect} & \textbf{$p < .05$} \\
        \hline
        (call, FM, FM)  & 48 & 67\\

        (call, FM, Accessor) & 13 & 33\\

        (contain, FM, Class) & 11 & 26\\
        
        (use, FM, Field) & 10 & 27\\
        
        (call, FM, Constructor) & 6 & 32\\
        
        (create, FM, Class) & 6 & 32\\
        
        (parameter, FM, Class) & 4 & 23\\
        
        (return, FM, Class) & 4 & 25\\

        (use, Field, Field) & 2 & 7\\
        \hline
    \end{tabular}
    \label{tab:dep_type_frequency}
\end{table}

Table~\ref{tab:dep_type_frequency} presents, for each dependency type, the number of tests rejecting $H_0$ with at least a small effect size; counts are zero for dependency types not shown. The table also reports the number of tests rejecting $H_0$ regardless of effect size.

For each dependency type, up to 80 tests were conducted (i.e., 10 code smell pairs $\times$ 4 contrasts $\times$ 2 flow directions = 80). However, it was not always possible to conduct all tests for every dependency type, as explained in Section~\ref{sec:data_analysis}. For reference, 72 tests were conducted for (call, FM, FM), and 40 for the other dependency types shown in the table. The dependency types most frequently rejecting $H_0$ with at least a small effect size are (call, FM, FM), (call, FM, Accessor), (contain, FM, Class), and (use, FM, Field), accounting for 25\%-67\% of all conducted tests. These proportions increase to 65\%-93\% when the small-effect-size threshold is removed.

The (call, FM, Accessor) and (use, FM, Field) dependencies indicate an important behavior involving data access. We therefore refer to these two dependency types as \textbf{\emph{data access}}. The (call, FM, FM) dependency type represents another key behavior, in which one FM depends on the functionality of another FM; we refer to this as an \textbf{\emph{FM call}}. Other dependencies, such as (contain, FM, Class), are likely used for reference purposes to support these two kinds of dependencies when artifacts are in different classes.

{%
\def\PadX{12pt}%
\def\PadY{4.5pt}%
\setlength{\fboxsep}{0pt}%

\def\FrameCommand#1{%
  {\color{Findingbar}\vrule width 2pt}%
  \colorbox{shadecolor}{%
    \hspace*{\PadX}%
    \vbox{\kern\PadY #1 \kern\PadY}%
    \hspace*{\PadX}%
  }%
}%
\MakeFramed{\advance\hsize-\width \FrameRestore}%
\small\noindent
\textbf{Finding 2.} Code smell interactions are frequently associated with significant differences in data access and FM call dependencies, observed in 25\%–67\% of tests with at least a small effect size, and up to 93\% when the small-effect-size threshold is removed.
\endMakeFramed
}

\begin{figure}[t]
\centering
  \includegraphics[width=\columnwidth]{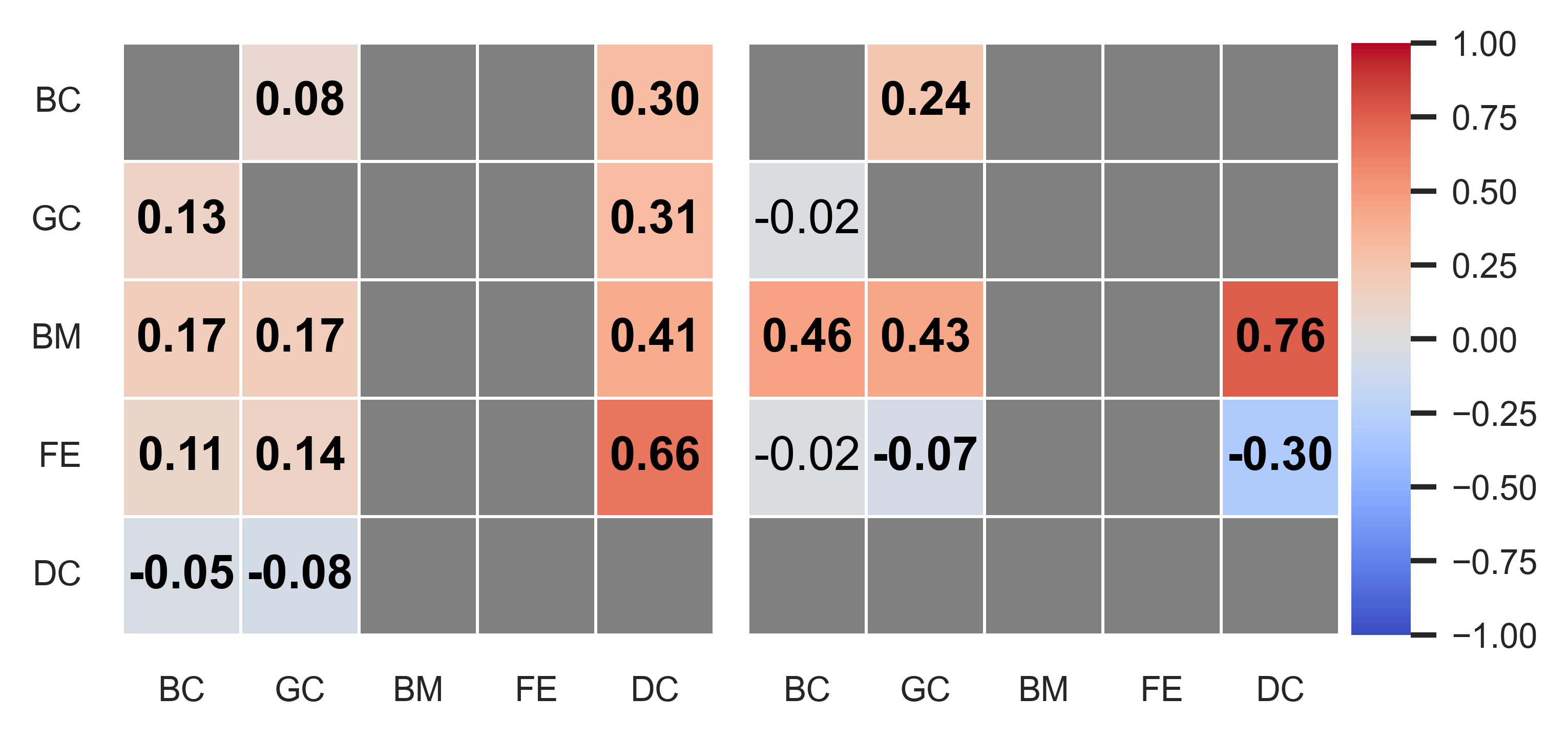}
  \caption{Cliff's delta: \textbf{CS1$\rightarrow$CS2} vs. \textbf{nonCS1$\rightarrow$CS2} (data access dependency)}~\label{fig:data_seeker}
  \Description{Cliff's delta: \textbf{CS1$\rightarrow$CS2} vs. \textbf{nonCS1$\rightarrow$CS2} (data access dependency)}
\end{figure}

\begin{figure}[t]
\centering
  \includegraphics[width=\columnwidth]{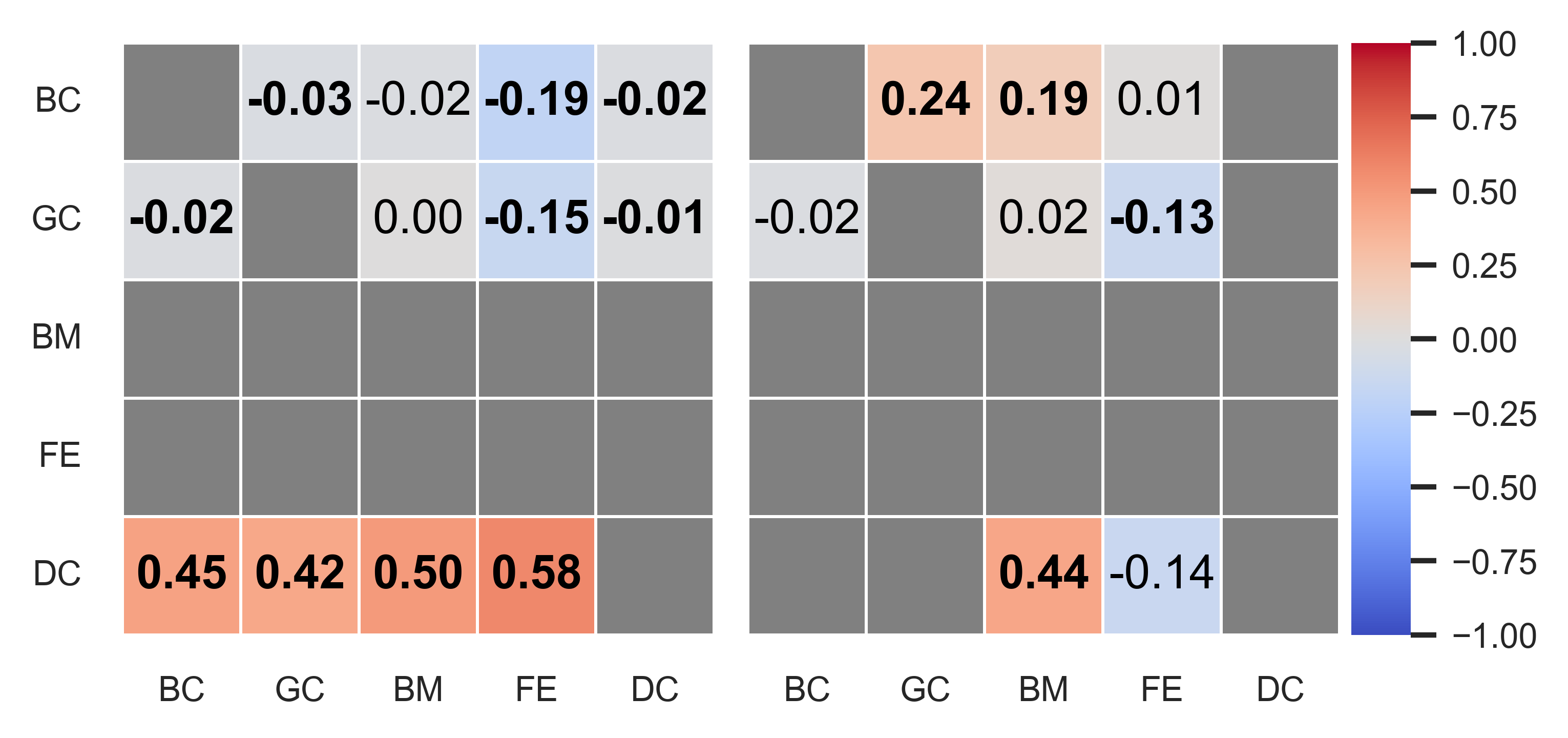}
  \caption{Cliff's delta: \textbf{CS1$\leftarrow$CS2} vs. \textbf{nonCS1$\leftarrow$CS2} (data access dependency)}~\label{fig:data_provider}
  \Description{Cliff's delta: \textbf{CS1$\leftarrow$CS2} vs. \textbf{nonCS1$\leftarrow$CS2} (data access dependency)}
\end{figure}

\begin{figure}[t]
\centering
  \includegraphics[width=\columnwidth]{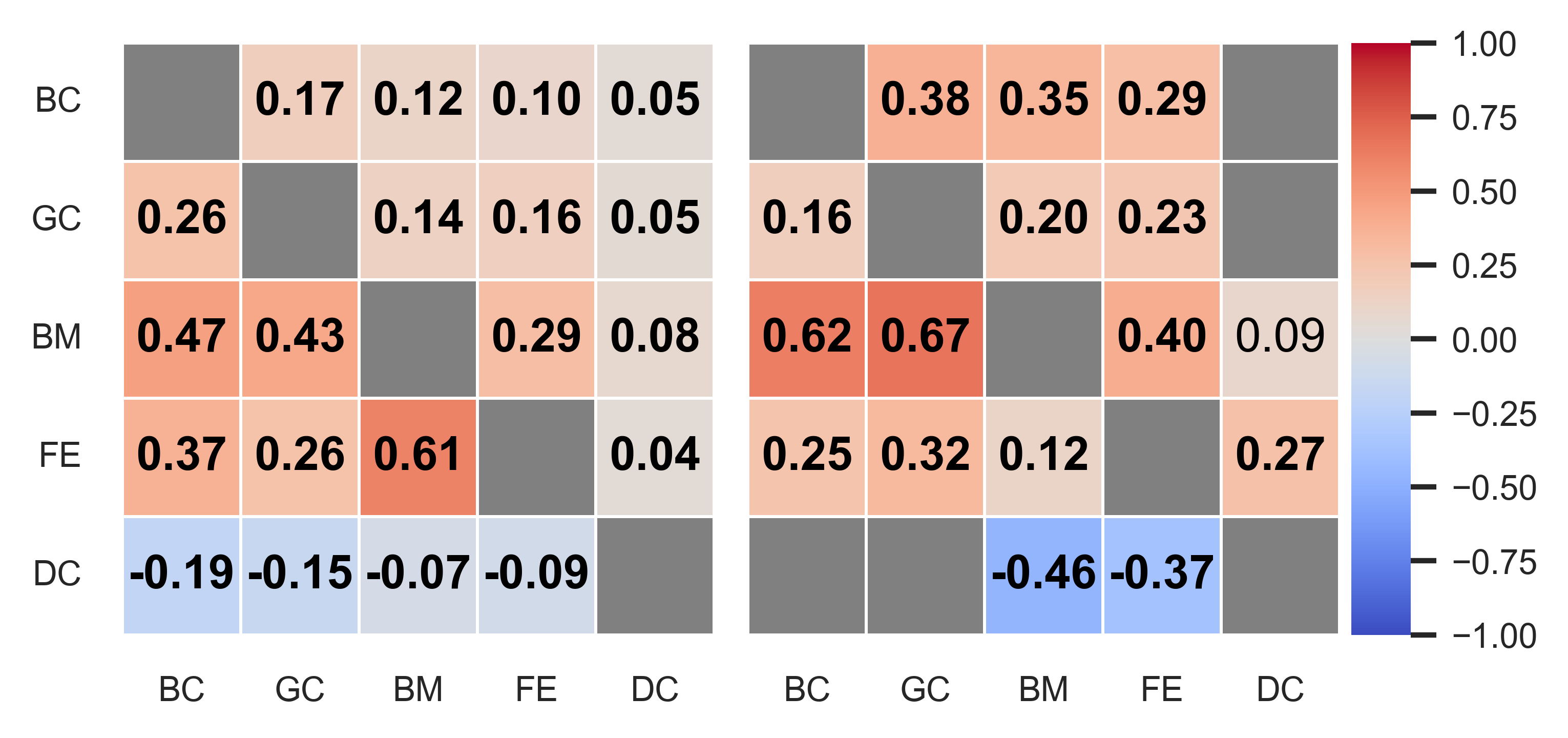}
  \caption{Cliff's delta: \textbf{CS1$\rightarrow$CS2} vs. \textbf{nonCS1$\rightarrow$CS2} (FM call dependency)}~\label{fig:FM_seeker}
  \Description{Cliff's delta: \textbf{CS1$\rightarrow$CS2} vs. \textbf{nonCS1$\rightarrow$CS2} (FM call dependency)}
\end{figure}

\begin{figure}[t]
\centering
  \includegraphics[width=\columnwidth]{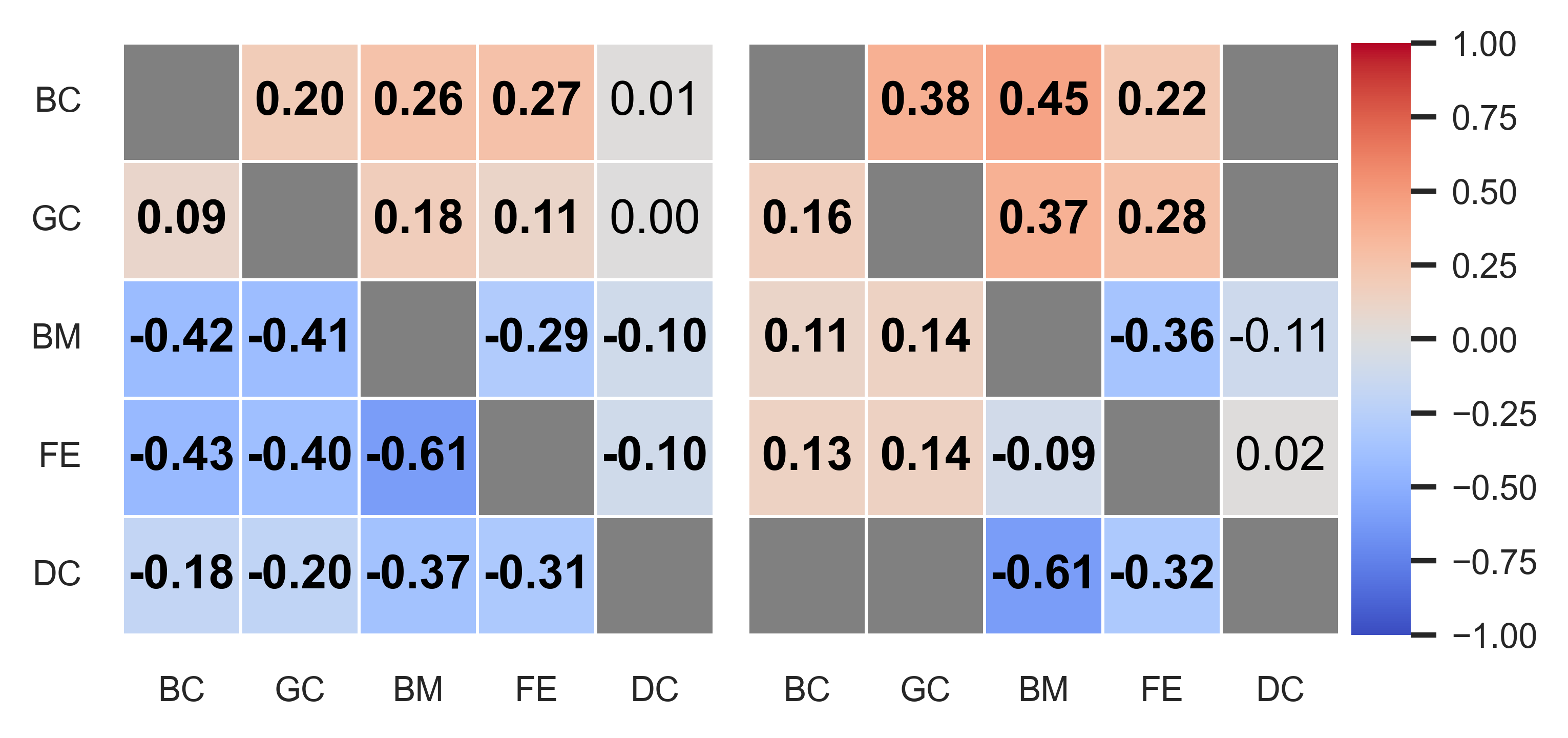}
  \caption{Cliff's delta: \textbf{CS1$\leftarrow$CS2} vs. \textbf{nonCS1$\leftarrow$CS2} (FM call dependency)}~\label{fig:FM_provider}
  \Description{Cliff's delta: \textbf{CS1$\leftarrow$CS2} vs. \textbf{nonCS1$\leftarrow$CS2} (FM call dependency)}
\end{figure}

For two code smells, CS1 and CS2, when we consider data access and FM call dependencies in a specific flow direction (e.g., \textbf{CS1$\rightarrow$CS2}), one artifact plays the \textbf{\emph{role}} of providing data or FM (e.g., the CS2 artifact), which we call the \textbf{\emph{provider}}; the other artifact plays the role of consuming the data or FM (e.g., the CS1 artifact), which we call the \textbf{\emph{consumer}}.

Figures \ref{fig:data_seeker} to \ref{fig:FM_provider} present several 5$\times$5 Cliff’s delta heatmaps. Each cell at row CS1 and column CS2 shows Cliff’s delta for the data access or FM call dependency of \textbf{CS1$\rightarrow$CS2} vs. \textbf{nonCS1$\rightarrow$CS2}, where CS1 and nonCS1 act as consumers and CS2 acts as the provider; or \textbf{CS1$\leftarrow$CS2} vs. \textbf{nonCS1$\leftarrow$CS2}, where CS1 and nonCS1 act as providers and CS2 acts as the consumer. Because the heatmap is a square matrix, both orders of each code smell pair are represented, so both flow directions of the four contrasts are covered for each dependency type. Figures \ref{fig:data_seeker} and \ref{fig:data_provider} cover the 40 tests for data access dependency, while Figures \ref{fig:FM_seeker} and \ref{fig:FM_provider} cover the 72 tests for FM call dependency. In each figure, the left heatmap represents the different-class situation, and the right heatmap represents the same-class situation. For FM call dependencies, we show (call, FM, FM). For data access dependencies, we show (call, FM, Accessor) for the different-class situation and (use, FM, Field) for the same-class situation, since cross-class data access is typically achieved through accessor methods, while within-class data access usually occurs through direct field use. Cells in bold indicate cases where we can reject $H_0$, and cells masked in gray represent cases that are impossible or do not exist in our dataset, as explained in Section~\ref{sec:data_analysis}.

We can reject $H_0$ for most tests. Within each row, most cells share the same change direction, particularly when the artifacts belong to different classes. This indicates that most code smells consistently show either an increase or a decrease in FM call and data access dependencies when interacting with another smell. However, the direction may differ by relative location. For example, in Figure \ref{fig:data_seeker}, the FE row shows an increase in data access dependencies in different-class scenarios but a decrease in same-class scenarios. This could be explained by the nature of FE, which tends to access data from other classes and is less concerned with data from its own class. In addition, the change direction may vary across dependency types and roles. For example, DC is associated with an increase in providing data to BC ($\delta = 0.45$, Figure~\ref{fig:data_provider}) but with a decrease in providing FM to BC ($\delta = -0.18$, Figure~\ref{fig:FM_provider}) and a decrease in consuming data from BC ($\delta = -0.05$, Figure~\ref{fig:data_seeker}).

{%
\def\PadX{12pt}%
\def\PadY{4.5pt}%
\setlength{\fboxsep}{0pt}%

\def\FrameCommand#1{%
  {\color{Findingbar}\vrule width 2pt}%
  \colorbox{shadecolor}{%
    \hspace*{\PadX}%
    \vbox{\kern\PadY #1 \kern\PadY}%
    \hspace*{\PadX}%
  }%
}%
\MakeFramed{\advance\hsize-\width \FrameRestore}%
\small\noindent
\textbf{Finding 3.} For a given relative location and role, most code smells show a consistent change direction in the number of data access and FM call dependencies when interacting with another code smell, although the direction varies across dependency types, locations, and roles.
\endMakeFramed
}

\subsection{Per-System General and Specific Analyses}

\begin{figure}[t]
\centering
  \includegraphics[width=\columnwidth]{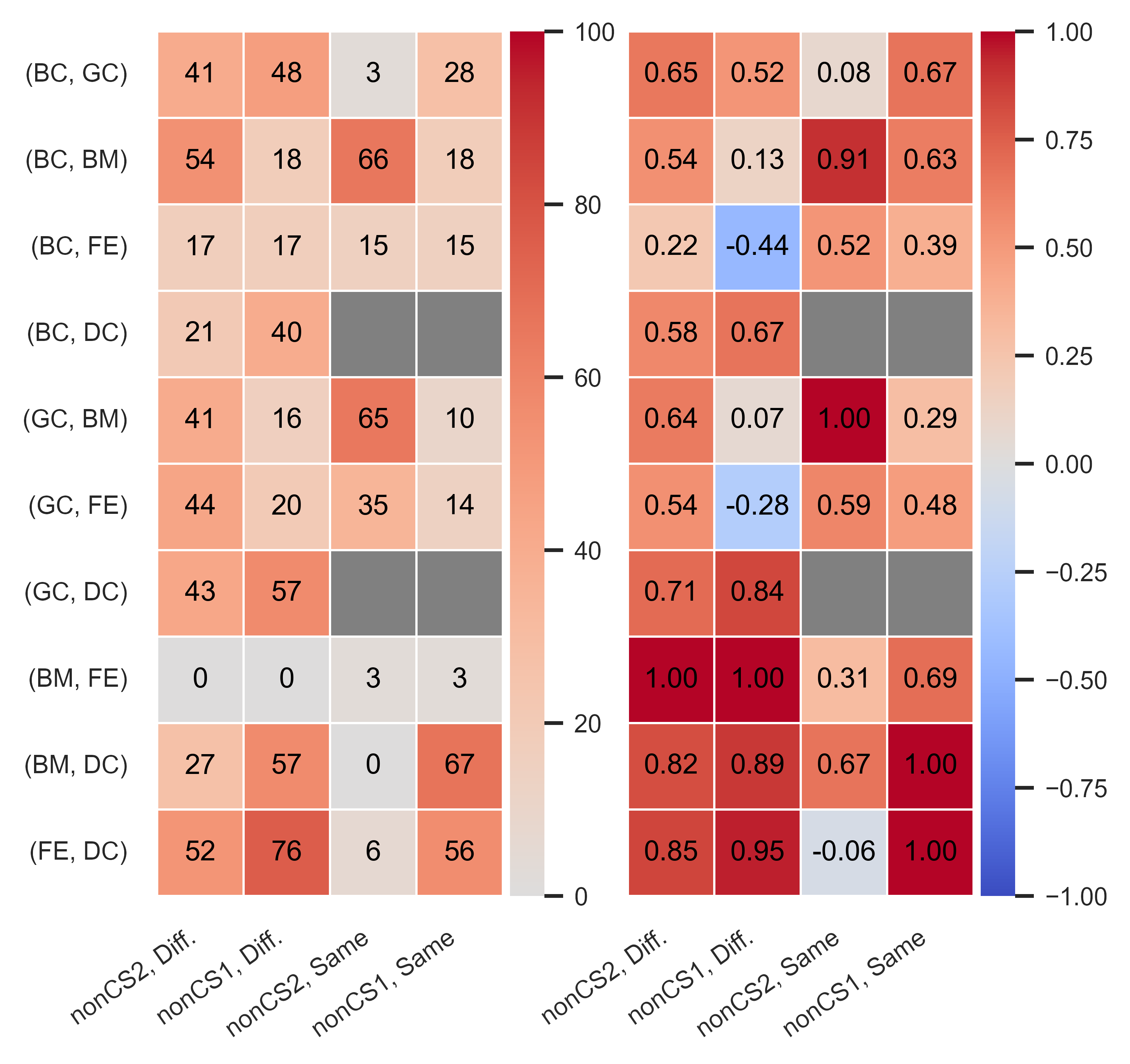}
  \caption{Significance rate and consistency score across all projects}~\label{fig:sig_rate_consistency}
  \Description{Significance rate and consistency score across all projects}
\end{figure}

Figure \ref{fig:sig_rate_consistency} presents two heatmaps for the per-system general analysis. We include only the projects for which the statistical test can be conducted, as discussed in Section \ref{sec:data_analysis}. The left heatmap shows the \textbf{\emph{significance rate}}, defined as the percentage of projects that yield a statistically significant result. The right heatmap shows the \textbf{\emph{consistency score}}, calculated as the difference between the number of projects with a nonnegative Cliff’s delta and those with a negative Cliff’s delta, divided by the total number of included projects. The score ranges from -1 to 1 and indicates whether projects tend to share the same change direction. A value near 0 indicates low consistency, whereas values near -1 or 1 indicate perfect consistency (all increases or all decreases, respectively). The significance rate ranges from 0\% to 76\%, and the consistency score ranges from -0.44 to 1.00, indicating that, for some code smell pairs, different projects exhibit mixed significance levels and opposite change directions.

\begin{figure}[tb]
\centering
  \includegraphics[width=0.95\columnwidth]{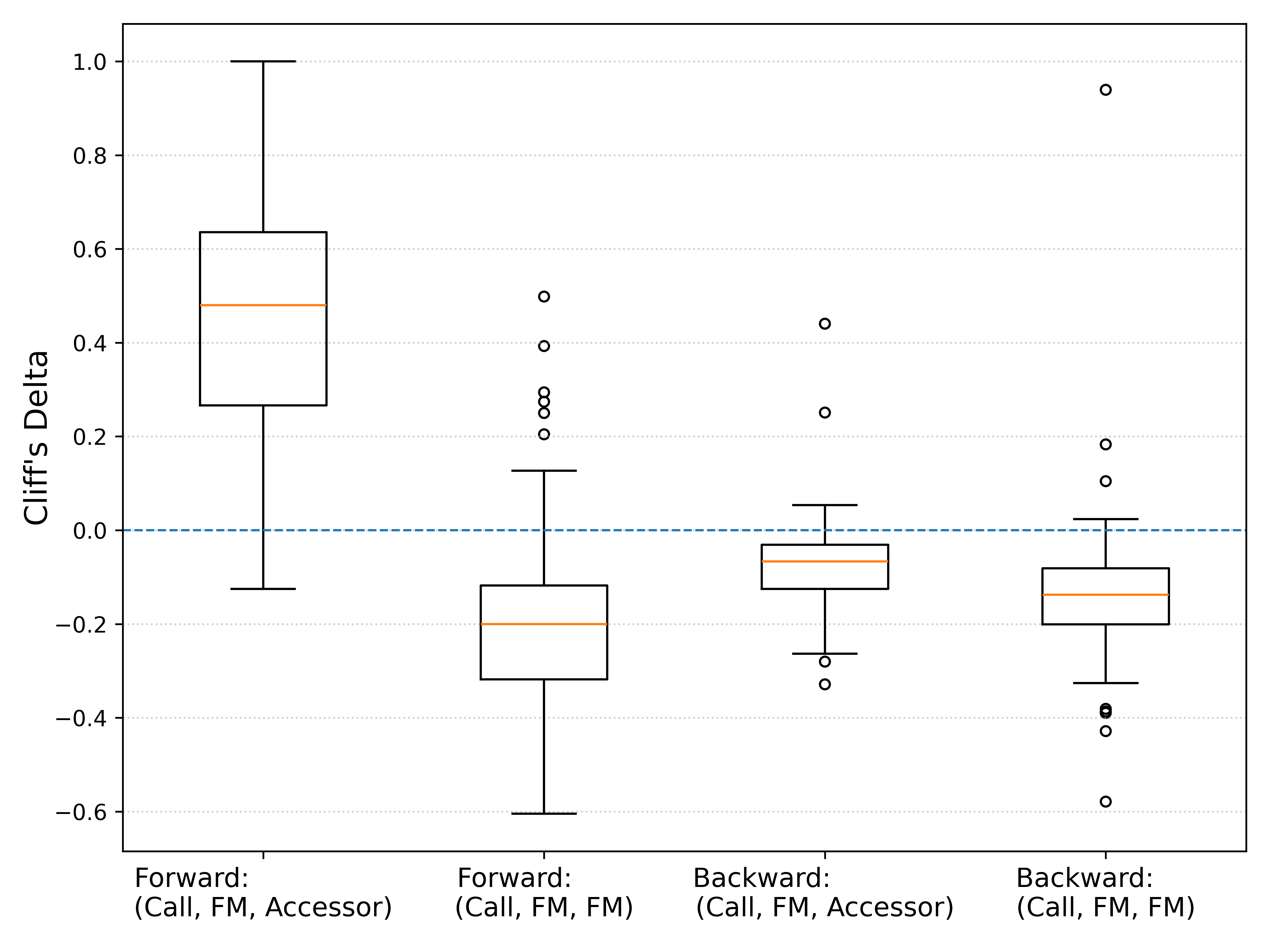}
  \caption{Cliff’s delta for FM call and data access dependencies for ``(GC–DC) vs. (GC–nonDC), Different'' across all projects, shown for both forward (GC$\rightarrow$DC vs. GC$\rightarrow$nonDC) and backward (DC$\rightarrow$GC vs. nonDC$\rightarrow$GC) flow directions}~\label{fig:boxplot}
  \Description{Cliff’s delta for FM call and data access dependencies for ``(GC–DC) vs. (GC–nonDC), Different'' across all projects, shown for both forward (GC$\rightarrow$DC vs. GC$\rightarrow$nonDC) and backward (DC$\rightarrow$GC vs. nonDC$\rightarrow$GC) flow directions}
\end{figure}

Consider the cell at row (GC, DC) and column (nonCS2, Diff.) in Figure~\ref{fig:sig_rate_consistency}, which compares GC-DC with GC-nonDC. It has a significance rate of $43\%$ and a consistency score of $0.71$. This indicates that about half of the projects are not significant, with a tendency toward a nonnegative Cliff’s delta. Figure~\ref{fig:boxplot} presents the corresponding per-system specific analysis for this pair, focusing on FM call and data access dependencies when the artifacts are in different classes. For all dependency types in both flow directions, over $75\%$ of individual projects follow the same change direction observed in the Union dataset (Figures~\ref{fig:data_seeker}-\ref{fig:FM_provider}). Most projects that do not follow this pattern are either outliers, yield non-significant results, or contain limited data that reduce the power of statistical tests. We also observe that Cliff’s delta for the forward (call, FM, Accessor) dependency tends to be positive, whereas the other three tend to be negative, with the change magnitude varying across projects. These opposite change directions among dependency types, along with the varying magnitudes across projects, help explain why, in the per-system general analysis, certain contrasts exhibit low significance rates and consistency across projects.

{%
\def\PadX{12pt}%
\def\PadY{4.5pt}%
\setlength{\fboxsep}{0pt}%

\def\FrameCommand#1{%
  {\color{Findingbar}\vrule width 2pt}%
  \colorbox{shadecolor}{%
    \hspace*{\PadX}%
    \vbox{\kern\PadY #1 \kern\PadY}%
    \hspace*{\PadX}%
  }%
}%
\MakeFramed{\advance\hsize-\width \FrameRestore}%
\small\noindent
\textbf{Finding 4.} Across projects, change magnitudes vary for both data access and FM call dependencies, yet the change direction remains largely consistent and aligns with the trend observed in the Union dataset.
\endMakeFramed
}

{%
\def\PadX{12pt}%
\def\PadY{4.5pt}%
\setlength{\fboxsep}{0pt}%

\def\FrameCommand#1{%
  {\color{RQbar}\vrule width 2pt}%
  \colorbox{shadecolor}{%
    \hspace*{\PadX}%
    \vbox{\kern\PadY #1 \kern\PadY}%
    \hspace*{\PadX}%
  }%
}%
\MakeFramed{\advance\hsize-\width \FrameRestore}%
\small\noindent
\textbf{RQ2 Summary.} Overall, code smell interactions are generally associated with an increase in the total number of dependencies. Significant changes with at least a small effect size are most frequently observed in data access and FM call dependencies. For a given relative location and role, individual code smells tend to exhibit a consistent change direction in these dependency types when interacting with other smells.
\endMakeFramed
}

\section{Discussion} \label{sec:discussion}

\subsection{A Multi-Dimensional Typology}

Since all code smells exhibit a consistent change direction in FM call and data access dependencies when interacting with other code smells under different class situations, we propose a \textit{multi-dimensional typology} as shown in Table~\ref{tab:typology}. Each dimension is defined by the dependency type (FM call or data access; for simplicity, we refer to them as FM and Data) and the dependency role (consumer or provider). Each dimension contains two categories: \textbf{\emph{strong}} and \textbf{\emph{weak}}. A code smell is classified as strong for a given dependency type and role if, compared to its non-code smell counterpart, it is associated with an increase in the number of dependencies when interacting with another smell. Conversely, it is classified as weak if it is associated with a decrease in the number of such dependencies.

This typology can be used to predict the direction of changes in data access and FM call dependencies during code smell interactions across different class situations. For example, we would expect increasing data access dependencies for FE$\rightarrow$DC vs. nonFE$\rightarrow$DC, since FE is a strong data consumer. Such insights can guide more targeted refactoring strategies and enhance both the identification and prioritization of code smells. Sections \ref{subsec:design_insghts}, \ref{subsec:interaction_based_inference}, and \ref{subsec:interaction_guided_refactoring} illustrate the potential practical utility of this typology.

\begin{table}[t]
\caption{A Multi-Dimensional Typology Based on Dependency Type and Role}
\Description{A Multi-Dimensional Typology Based on Dependency Type and Role}
\label{tab:typology}
\centering
\small
\begin{tabular}{|l|cc|cc|}
\hline
\multirow{2}{*}{\textbf{Smell}} & \multicolumn{2}{c|}{\textbf{Data}} & \multicolumn{2}{c|}{\textbf{FM}} \\
\cline{2-5}
 & \multicolumn{1}{c|}{\textbf{Provider}} & \textbf{Consumer} & \multicolumn{1}{c|}{\textbf{Provider}} & \textbf{Consumer} \\
\hline
DC & S & W & W & W \\
GC & W & S & S & S \\
BC & W & S & S & S \\
BM & NA & S & W & S \\
FE & NA & S & W & S \\
\hline
\multicolumn{5}{l}{S = Strong; W = Weak; NA = Not Applicable} \\
\end{tabular}
\end{table}

\subsection{Design Insights Derived from Dependency Distribution}
\label{subsec:design_insghts}

FE and BM are weak FM providers, meaning their functionality is less likely to be relied upon during interactions compared to non-code smell artifacts. Therefore, it is recommended to create small methods, not only because large ones are harder to understand, but also because as methods grow, they often take on multiple responsibilities that are not needed simultaneously, reducing their usefulness as single-responsibility units for other artifacts. In contrast, GC and BC are strong FM providers and consumers, invoking and offering many FMs due to their large and multi-responsibility nature. Therefore, creating small, single-responsibility classes is advisable.

\subsection{Interaction-Based Inference of Code Smells}
\label{subsec:interaction_based_inference}

It has been suggested that the presence of certain types of smells can support the existence of other types of smells \cite{pietrzak2006leveraging} (e.g., the existence of data classes supports the existence of feature envy methods that use their data). Such information can improve the efficiency and effectiveness of detecting additional code smells when certain code smells have been confirmed (e.g., manually confirmed). We can also leverage interaction information and the typology to offer another perspective on evidence of support.

Given a confirmed code smell in one artifact, we assess whether a related artifact exhibits a different smell by examining their interactions. Specifically, if the expected change direction derived from the typology aligns with the observed change direction in the number of dependencies, we can increase our confidence that the related artifact exhibits the code smell.

For example, consider a GC that has been manually confirmed. Suppose there is a method $m_1$ for which we want to detect whether it exhibits FE. If $m_1$ does not interact with the GC, then we cannot utilize interaction information. However, if they do interact, one useful interaction signal is the number of FM call dependencies from the GC to each method it interacts with, including $m_1$. If the number of FM calls from the GC to $m_1$ is relatively low compared to those to other methods that the GC interacts with, this provides supporting evidence that $m_1$ exhibits FE. Since a GC is a strong FM consumer, we would typically expect a high number of FM calls when it interacts with other artifacts. However, if $m_1$ exhibits FE, then it is a weak FM provider and will be called less often (i.e., it will receive fewer FM calls) compared to nonFE methods that the GC interacts with. Thus, a low number of FM calls from the GC to $m_1$ supports the existence of FE in $m_1$.

Furthermore, interactions involving additional confirmed code smells can further strengthen our confidence. For instance, if a confirmed DC also interacts with $m_1$, a relatively high number of data access dependencies from $m_1$ to the DC, compared with those from other methods that interact with the DC, would further support that $m_1$ exhibits FE. If $m_1$ exhibits FE, then it is a strong data consumer and the DC is a strong data provider; therefore, such interactions are expected to exhibit a high number of data access dependencies.

\subsection{Interaction-Guided Refactoring Strategies}
\label{subsec:interaction_guided_refactoring}

Developers should prioritize interacting code smells for two reasons: (i) they typically involve more dependencies, making the code harder to understand and more error-prone to change. Fixing an interacting pair collapses a larger bundle of edges and reduces ripple effects; and (ii) unlike raw coupling, which mixes intentional and harmful edges, interactions reveal where dependency growth is structurally risky because it concentrates in specific dependency types and directions. These patterns usually indicate misplaced responsibilities and broken encapsulation. Combined with the characteristics of the smells involved, this makes fixing more actionable, providing clear refactoring strategies that can address multiple smells at once, compared with equally coupled modules without smell interactions.

For example, when FE interacts with DC, we expect a higher-than-usual number of data access dependencies from FE to DC. This suggests a potential refactoring strategy: move FE into DC. Doing so would prevent FE from envying DC’s data and would add a functional method to DC, making it less of a pure data holder. In Grasscutter\footnote{Grasscutter is an open-source anime game server emulator. Source code: https://github.com/Grasscutters/Grasscutter}, \textit{Friendship\#toProto} (an FE) heavily accesses the data of \textit{PlayerProfile} (a DC) to serialize it, suggesting that \textit{Friendship\#toProto} could be moved into \textit{PlayerProfile} to reduce foreign data access and make the DC more behaviorally rich.

Another example involves GC interacting with BC, where we observe a higher-than-usual number of data access and FM call dependencies. Both BC and GC represent large, complex classes that often contain multiple responsibilities. A common refactoring strategy is to extract classes from them, each representing a single responsibility \cite{lacerda2020code}. However, the extensive dependencies between them suggest that the responsibilities may be interwoven across the two classes. Therefore, instead of extracting separate classes from each, it may be better to extract a shared class from both, considering their interactions. In Grasscutter, methods such as \textit{upgradeWeapon}, \textit{promoteWeapon}, and \textit{calcWeaponUpgradeReturnItems} in \textit{InventorySystem} (a BC) repeatedly operate on the same \textit{GameItem}'s (a GC) fields, such as \textit{level} and \textit{exp}, and always conclude with a call to \textit{save}. Extracting these methods and the related data into a dedicated \textit{WeaponProgressionService} would create a cohesive unit focused on weapon advancement, reducing foreign data access and coupling, and decreasing the size and complexity of both original classes.

\subsection{Threats to Validity}
\label{subsec:threat}

As with any similar study, there is a threat to external validity due to the systems we studied. Our conclusions may still apply to languages similar to Java, such as C++ and C\#. It would be extremely interesting if the language were significant. The expensive nature of our analysis limited the sample size. A post-hoc analysis revealed that the selected projects vary in size, number of code smells, and domains. Based on this, we concluded that our sample is sufficiently representative of Java projects. While there is no reason to expect a larger sample would change the conclusions, we do note that some individual projects did not follow the trend we observed.

The choice of detection tool represents a potential threat to construct validity. Many existing tools lack clear documentation of their detection heuristics and code smell definitions, and they often support only a limited subset of smells. To address this, we developed our own tool based on the metric-based detection framework proposed by Lanza and Marinescu \cite{lanza2007object} and made it publicly available to enhance transparency and replicability.

A threat to internal validity arises from our use of stricter detection thresholds to minimize false positive rates; as a consequence, false negative rates may increase. To assess this threat, we conducted a manual evaluation, which indicates that both false positive and false negative rates are low, although false negative rates are slightly higher. Undetected smell instances would be misclassified into non-smelly groups, diluting the contrast between interaction types and biasing effect sizes toward zero. However, given the overall low false positive and false negative rates, we expect this threat to be minimal.

Another threat to internal validity arises from how we define non-code smell artifacts: a nonCS1 artifact may still exhibit other smells (e.g., CS2), which could potentially introduce confounding effects. To assess this threat, we conducted a sensitivity analysis by excluding nonCS artifacts that exhibit any of the other four smells and reran the analysis. The results show that such contamination is rare and that our main findings remain consistent.

A further threat to internal validity is that inherently different dependency levels between code smell and non-code smell artifacts may confound our comparisons; that is, the observed differences in dependency distributions may be caused by differences in artifact dependency levels rather than the interaction itself. Since our study is observational, we do not claim causal effects. We interpret the results as associations between smell interactions and dependency distributions.

\section{Conclusions and Future Work} \label{sec:conclusions}

The objective of this study is to investigate the frequency of code smell interactions and how dependency distribution varies due to these interactions. This is achieved by examining five code smells across 116 open-source Java projects.

Our results suggest that most pairs of code smells interact frequently, and that these interactions are associated with a significant increase in the total number of dependencies. Although this outcome may not be unexpected, to the best of our knowledge, this study provides the first rigorous empirical evidence to substantiate this relationship. Moreover, code smell interactions exhibit consistent change direction in certain dependency types. These findings offer meaningful design insights, suggesting that practitioners should prioritize resolving interacting code smells rather than those occurring in isolation. Interacting smells not only display denser and more complex dependency structures but also show consistent patterns of dependency change that may reflect deeper design issues. At the same time, this information can guide more effective refactoring strategies and inform the development of improved code smell detection techniques.

Future work could investigate a broader range of code smells, interactions among code smells of the same type, and the influence of these interactions on other maintainability metrics. Additionally, contextual factors such as project domain and programming language could be examined to assess their impact on code smell interactions. Furthermore, interaction effects could be better isolated to determine whether smell interactions have causal impacts on dependency distributions.

Future work could also formulate and empirically test hypotheses that link dependency distribution patterns to process outcomes (e.g., change proneness, churn, or effort). While this study does not directly evaluate process outcomes, our results on dependency distribution patterns provide a foundational characterization that can inform such analyses. For example, increased dependencies between interacting smelly artifacts may create additional pathways for change propagation and thus be associated with higher change proneness and maintenance effort.

\bibliographystyle{ACM-Reference-Format}
\bibliography{sections/References}

\end{document}